\documentclass[11pt]{article}
\usepackage{amssymb}\usepackage{amsfonts}
\usepackage{amsmath} 
\renewcommand{\arraystretch}{1.6}

   \renewcommand{\theequation}{\thesection.\arabic{equation}}
   \csname @addtoreset\endcsname{equation}{section}
    \csname @addtoreset\endcsname{figure}{section}
    \csname @addtoreset\endcsname{table}{section}

\newcounter{thanksnum}
\def\thanksnumber#1
{\setcounter{thanksnum}{\value{footnote}}\setcounter{footnote}{#1}%
                     \addtocounter{footnote}{-1}\footnotemark
                     \setcounter{footnote}{\value{thanksnum}}}
\def\newtheoremz#1{\@ifnextchar[{\@othmz{#1}}{\@nthmz{#1}}}

\def\@nthmz#1#2{%
\@ifnextchar[{\@xnthmz{#1}{#2}}{\@ynthmz{#1}{#2}}}

\def\@xnthmz#1#2[#3]{\expandafter\@ifdefinable\csname #1\endcsname
{\@definecounter{#1}\@addtoreset{#1}{#3}%
\expandafter\xdef\csname the#1\endcsname{\expandafter\noexpand
  \csname the#3\endcsname \@thmcountersepz \@thmcounterz{#1}}%
\global\@namedef{#1}{\@thmz{#1}{#2}}\global\@namedef{end#1}{\@endtheoremz}}}

\def\@ynthmz#1#2{\expandafter\@ifdefinable\csname #1\endcsname
{\@definecounter{#1}%
\expandafter\xdef\csname the#1\endcsname{\@thmcounterz{#1}}%
\global\@namedef{#1}{\@thm{#1}{#2}}\global\@namedef{end#1}{\@endtheoremz}}}

\def\@othmz#1[#2]#3{\expandafter\@ifdefinable\csname #1\endcsname
  {\global\@namedef{the#1}{\@nameuse{the#2}}%
\global\@namedef{#1}{\@thmz{#2}{#3}}%
\global\@namedef{end#1}{\@endtheoremz}}}

\def\@thmz#1#2{\refstepcounter
    {#1}\@ifnextchar[{\@ythmz{#1}{#2}}{\@xthmz{#1}{#2}}}

\def\@xthmz#1#2{\@begintheoremz{#2}{\csname the#1\endcsname}\ignorespaces}
\def\@ythmz#1#2[#3]{\@opargbegintheoremz{#2}{\csname
       the#1\endcsname}{#3}\ignorespaces}

\def\@thmcounterz#1{\noexpand\arabic{#1}}
\def\@thmcountersepz{.}
\def\@begintheoremz#1#2{ \trivlist \item[\hskip \labelsep{\bf #1\ #2}]}
\def\@opargbegintheoremz#1#2#3{ \trivlist
      \item[\hskip \labelsep{\bf #1\ #2\ (#3)}]}
\def\@endtheoremz{\endtrivlist}

\newtheorem{theorem}{Theorem}[section]
\newtheorem{lemma}{Lemma}[section]

\newtheorem{proposition}{Proposition}[section]

\newtheorem{definition}{Definition}[section]
\newtheorem{remark}{Remark}[section]

\newtheorem{example}{Example}[section]

\def\e{\varepsilon}

\def\defi{\stackrel{{\scriptscriptstyle \Delta}}{=}}

\def\a{\alpha}

\def\o{\omega}
\def\O{\Omega}

\def\Y{{\cal Y}}
\def\F{{\cal F}}
\def\w{\widehat}
\def\Ind{{\,\rm Ind\,}}
\def\Ind{{\mathbb{I}}}

\def\sign{{\rm  sign\,}}

\def\esssup{\mathop{\rm ess\, sup}}
\def\essinf{\mathop{\rm ess\, inf}}

\def\R{{\bf R}}
\def\E{{\bf E}}
\def\P{{\bf P}}

\def\Z{{\cal Z}}

\def\H{{\cal H}}

\def\L{L}

\def\b{\beta}
\def\s{\delta}
\def\g{\gamma}

\def\ww{\widetilde}
\def\X{{\cal X}}
\def\t{\theta}
\def\oo{\bar}
\def\s{\sigma}

\def\V{{\cal V}}

\def\M{{\cal M}}

\def\L{{\cal L}}

\newcommand{\be}{\begin{equation}}
\newcommand{\ee}{\end{equation}}
\newcommand{\bd}{\begin{displaymath}}
\newcommand{\ed}{\end{displaymath}}
\newcommand{\ba}{\begin{array}{ll}}
\newcommand{\ea}{\end{array}}
\newcommand{\baa}{\begin{eqnarray}}
\newcommand{\eaa}{\end{eqnarray}}
\newcommand{\baaa}{\begin{eqnarray*}}
\newcommand{\eaaa}{\end{eqnarray*}}
\font\sm=cmr10

\def\oo{\bar}

\def\mar{{{\scriptscriptstyle market}}}
\def\imp{{{\scriptscriptstyle implied}}}

\def\FF{{\mathbb{F}}}

\def\QQ{{ Q}}

\def\T{{\cal T}}
\def\QQ{\P_\t }
\def\X{{\cal X}_\t }
\def\Y{{\cal Y}_\t }
\def\nuu{\nu}
\date{Submitted: August 3, 2011. Revised: September 30, 2014} 
\title{
On martingale measures and pricing for continuous bond-stock market with stochastic bond
\index{Earlier version of this paper was submitted
 on xxx.lanl.gov on August 3, 2011 under the tittle "On martingale measures and pricing for continuous bond-stock market with stochastic bond"}
 \index{On continuous bond-stock market with stochastic
deviations in bond prices}
}
\author{
Nikolai Dokuchaev \index{Corresponding address: Department of
Mathematics \& Statistics, Curtin University, GPO Box U1987, Perth,
6845 Western Australia,  email N.Dokuchaev@curtin.edu.au, tel. 61 8
92663144.} \\  {\sm Department of
Mathematics \& Statistics, Curtin University,}\vspace{-0.5cm} \\
\vspace{-0.5cm} {\sm  GPO Box U1987, Perth, 6845 Western Australia}}
\begin{document}
\maketitle
\begin{abstract}
This papers addresses the  stock option pricing problem in a  continuous
time  market model where there are two stochastic  tradable assets,
and one of them is selected as a num\'eraire. It is shown that the presence of  arbitrarily small
stochastic deviations in the evolution of the num\'eraire process
causes significant changes in the market properties. In particular,
an equivalent martingale measure is not unique for this market, and there are
non-replicable claims. The martingale prices and  the hedging error
can vary significantly and take extreme values, for some extreme
choices of the  equivalent martingale measures. Some rational choices of the  equivalent martingale
measures are suggested and discussed, including implied  measures  calculated from
observed bond prices. This allows to calculate the implied market price of risk process.
\end{abstract}
\section{Introduction} This paper addresses the  stock option pricing problem in a  continuous
time  market model. We consider the case where  there are two
stochastic tradable assets, and one of them is selected as a
num\'eraire.
This setting corresponds to a generalized  Black-Scholes model where stochastic deviations in the bond prices are allowed.
The classical Black and Scholes model includes a bond or money market account with the price $B(t)$  and a single risky asset with the
price $S(t)$; the process
$B(t)$ is assumed to be non-random or risk-free and is used as a num\'eraire, and $S(t)$ is assumed to be a stochastic
It\^o process with constant  volatility.  This  is a so-called complete market where any claim can be replicated
and where there is a unique  equivalent martingale (risk-neutral)  measure equivalent to the historical measure; the price of a derivative
is defined via the expectation of the discounted payoff by this unique  equivalent martingale  measure.

The pricing of derivatives is usually more difficult for so-called incomplete market models where
an  equivalent martingale measure is not unique; see, e.g.,
 El Karoui and Quenez  (1995).
 \par
 Some important examples of market incompleteness arise when the money market account is assumed to have a martingale part.
         Cheng (1991),  Kim
and Kunitomo (1999), Benninga {\em et al.} (2002), and Back (2010),
considered an  incomplete modification of the Black-Scholes model
where  $B(t)$ was an It\^o process. The cited papers considered
martingale pricing method where the option  price is calculated as
the expectation of the discounted claim under some  equivalent risk-neutral
measure (martingale measure) such that the discounted stock price
$S(t)/B(t)$ is a martingale on a given time interval $[0,T]$ under
this measure.  Cheng (1991) analyzed only one  equivalent martingale measure
among all  equivalent martingale measures; see Example \ref{exCheng} (iv).
Benninga {\em et al.} (2002) considered a multi-stock market under
requirements that make the choice of an  equivalent martingale measure unique
in the case of a single stock and stochastic bond; see Example
\ref{exCheng} (iii). Geman {\em et al.} (1995) considered pricing of
replicable claims only.  Kim and Kunitomo (1999) studied asymptotic
properties of this price with respect to a particular  equivalent martingale
measure. In the cited works, the pricing was studied for certain selected measures, and
the impact of non-uniqueness of an
 equivalent martingale measure was not discussed, as well as the presence of
non-replicable claims.
\par
For portfolio selection problems, related questions arise in the setting with
a random num\'eraire; see, e.g.,  Karatzas  and Kardaras (2007),
Kardaras (2010), Becherer  (2010).
\par
 In this paper, we revisited
the pricing problem for options for a market with two tradable stochastic assets.
 We consider a model with a continuum of
different  equivalent martingale measures and with claims
that cannot be replicated, even when the appreciation rate and
volatility coefficients  are constant
(Proposition \ref{ThM}).  We investigate  the impact
of  the presence of different equivalent martingale measures and
non-replicable claims. Our model  is
close to the model from Cheng (1991) with a modification that
ensures the existence of many equivalent martingale measures. Cheng
(1991) studied an impact of the  absence of an  equivalent martingale measure in
the setting  with a  stochastic bond price $B(t)$ such that $B(T)=1$. In
this case,  the appreciation rate of the
discounted stock price is imploding when terminal time is
approached, and  the Novikov's condition of existence of an equivalent martingale
measure is not satisfied. Our setting removes this feature; we
consider a stochastic num\'eraire  without restrictions on the terminal
price. This could be close to the model from Cheng (1991) if one considers
a stochastic bond  with the price $B(t)$ maturing at $T+\e$, i.e., such that $B(T+\e)=1$, for an
arbitrarily small $\e>0$.  We suggest some criterions of replicability (Theorems \ref{ThM00}  and \ref{ThM0}).  We investigated pricing of zero coupon bonds for this market,  and derived an equation for the associated risk-free rate (equations (\ref{bond})-(\ref{r}) in Section \ref{SecImp} and \ref{SecImp2}). Using these results, we obtained that, for a market model with stochastic num\'eraire, it is possible to estimate the market price of risk
and therefore the appreciation rate of the  stock price process as an implied parameter inferred from a single observed market  price of a zero coupon bond.
This is a novell result, since the existing literature does not consider implied appreciation rate; see discussion and citations in   Section \ref{SecImp}.
\par
In addition, we found that there are interesting features that makes
this model different from the models  with non-random num\'eraire and with the incompleteness caused by the randomness of the volatility. Apparently, arbitrarily small
deviations in the num\'eraire coefficients  cause significant changes in the market properties.  With these
deviations, the
martingale prices and the second moments of the hedging error
vary significantly and take extreme values for some extreme choices
of  equivalent martingale measures that represent  "extreme" beliefs of market agents (Theorems \ref{ThR1}-\ref{ThRelii}).
 This  can be interpreted as the following: among
the variety of  equivalent martingale measures, there are some measures that represent
extreme beliefs and can be excluded from the analysis.  We suggest a
parametrisation of the set of equivalent martingale measures that helps to select
reasonable equivalent martingale
measures, including a measure implied form observed prices of zero coupon bonds
(Theorems \ref{ThN} ). We discuss some particular choices of these measures, including a measure implied form observed prices of zero coupon bonds and   a measure that ensures local risk
minimizing hedging strategy (Examples \ref{Ex1}-\ref{Ex4} and Theorem \ref{ThLRM}).
\par
The rest of the paper is organized as follows.  In Section \ref{secSetting},  we describe the model setting.  In Section \ref{SecRep},  we discuss replicability of claims and the hedging errors.  In Section \ref{secE},  we show that the prices can take extreme values for some choices of martingale measures. In Section \ref{SecImp}, we suggest a method of calculation
the implied market price of risk for a case of complete market and review some related literature. In Section \ref{secS},  we discuss selection of the equivalent martingale measures.  In Section \ref{SecImp2}, we discuss
calculation of
the implied market price of risk for a more general case of an incomplete  market.
In Section \ref{secM}, parabolic equations are derived for the price, for the
hedging strategy, and for the hedging error, in a
Markovian setting.  The proofs are in the Appendix.
\section{Model setting}\label{secSetting}
We consider the diffusion model of a securities market  consisting
 of two tradable assets with the prices  $S(t)$ and $B(t)$,  ${t\ge 0}$. The prices evolve
 as \baaa \label{S} dS(t)=S(t)\bigl(a(t)dt+\s(t) dw(t)+\w\s(t)d\w w(t)\bigr),\quad
t>0, \eaaa and \baa \label{B} dB(t)=B(t)\bigl(\a(t) dt+\rho(t)
dw(t)+\w\rho(t)d\w w(t)\bigr).\eaa We assume that $W(t)=(w(t),\w w(t))$
is a standard Wiener process with independent components on  a given
standard probability space $(\O,\F,\P)$, where $\O$ is a set of
elementary events,
$\P$ is a probability measure, and  $\F$ is a $\P$-complete $\s$-algebra of events. The initial prices $S(0)>0$ and
$B(0)>0$ are given constants.
\par
We consider this model as an extension  of the classical bond and
stock market model, where a  bond with the price $B(t)$ is used as a num\'eraire, and its price evolves as    \baa \label{Bd}
\frac{dB}{dt}(t)=\a(t)B(t).\eaa
Equation (\ref{B}) for stochastic num\'eraire is a  generalization of (\ref{Bd}); one may say
that equation (\ref{B}) represents a modification  of equation for a
risk-free asset that takes into account possibility of stochastic
disturbances in the return rate. In this setting, $B(t)$ is not
exactly a risk-free asset. However, if the processes $\rho(t)$ and
$\w\rho(t)$ are small in some norm, then  (\ref{B}) can be
considered as the equation for the money market account  with small
deviations (see an example in Section \ref{SecRep} below).  In
particular, the conditions on the coefficients imposed below allow
the case where $\rho(t)\equiv \w\rho(t)=\e$ for an arbitrarily small
$\e>0$. We show below  that the presence
of arbitrarily small deviations in (\ref{B}) changes dramatically
the properties of the market model (see Section \ref{secE}).
\par
If $\esssup_{t,\o}(|\w\s(t,\o)|+|\w\rho(t,\o)|)> 0$,  we denote by $\FF=(\F_t)_{t\ge 0}$ the filtration generated by
the process $W=(w,\w w)$. If $\esssup_{t,\o}(|\w\s(t,\o)|+|\w\rho(t,\o)|)=0$,  we denote by $\FF=(\F_t)_{t\ge 0}$ the
filtration generated by the process $w$ only. In both cases, $\F_0$
is trivial, i.e., it is the $\P$-augmentation of the set
$\{\emptyset,\O\}$.
\par
We assume that the process
$\mu(t)=(a(t),\s(t),\w\s(t),\a(t),\rho(t),\w\rho(t))$ is $\FF$-adapted  and bounded.

Let $\ww\s=\s-\rho$ and  $\ww\rho=\w\s-\w\rho$. We assume that there exists $c>0$ such that either $|\ww\s(t,\o)|\ge c$ a.e. or $|\ww\rho(t,\o)|>c$ a.e.

\subsection*{Discounted stock price and  equivalent martingale measures}
Let $\ww S(t)\defi S(t)/B(t)$. By It\^o formula, it follows that this
process evolves as \baa &&d\ww S(t)=\ww S(t)\bigl(\ww a(t) dt+\ww\s(t)
dw(t)+\ww \rho(t) d\w w(t)\bigr),\\ &&\ww S(0)=S(0)/B(0),\label{wS}\eaa where
 \baaa   \ww a\defi a-\a+\rho^2 +\w\rho^2 -\s\rho -\w\s\w \rho.\eaaa
\par
Let  $V(t)=(V_1(t),V_2(t))^\top=(\ww\s(t),\ww\rho(t))^\top$ and  $\w
V(t)=(\w V_1(t),\w V_2(t))^\top=(\rho(t),\w\rho(t))^\top$. These
processes take values in $\R^2$. By the assumptions, $\essinf_{t,\o}|V(t,\o)|>0$.
\begin{definition}
Let $\T$ be the set of bounded $\FF$-adapted processes
$\t(t)=(\t_1(t),\t_2(t))^\top$ with values in $\R^2$ such that $
\t_1(t)\ww\s(t)+\t_2(t)\ww\rho(t)=\ww a(t), $ i.e., $V(t)^\top
\t(t)=\ww a(t)$.
\end{definition}

\par
For $\t\in\T$, set
 \be\label{ZT-1}
\Z_\t=\exp\left(-\int_0^T \t(s)^\top dW(s)- \frac{1}{2}\int_0^T
|\t(s)|^2ds\right). \ee Our standing assumptions imply that $\E
\Z_\t=1$.  Define the  probability measure $\P_\t $ by
$d\P_\t/d\P=\Z_\t$; this measure is equivalent to the measure $\P$.
Let $\E_\t$ be the corresponding expectation.

 Let
\renewcommand{\arraystretch}{1.0}\baa W_\t(t)=\left(
            \begin{array}{c}
              W_{\t 1}(t) \\
               W_{\t 2}(t) \\
            \end{array}
          \right)=
  \int_0^t\t(s)ds +W(t).\label{W}
\eaa \renewcommand{\arraystretch}{1.4} By Girsanov's Theorem, $W_\t$
is a standard Wiener process in $\R^2$ under $\P_\t$.

For $\t\in\T$, equation (\ref{wS}) can be rewritten as
\baaa &&d\ww S(t)=\ww S(t)V(t)^\top dW_\t(t).\eaaa
\begin{remark} Clearly, the set $\T$ has more than one element; it is
a linear manifold. Therefore, the selection of the process $\t(t)$
and the measure  $\P_\t$,  is not unique.
\end{remark}

\begin{example} \label{exCheng} {\rm \begin{itemize}
\item[(i)] If $\ww\rho\equiv 0$, then the process $\t_1(t)$ is uniquely
defined as
  $\t_1(t)=\ww\s(t)^{-1}\ww a(t)$.  If, in addition, the process $\ww\s(t)$ is non-random, then the process $\ww
  S(t)$ has the same distribution under $\P_\t$ for all $\t\in\T$.
\item[(ii)]
  If $\ww\s\equiv 0$, then the process $\t_2(t)$ is uniquely
defined as
  $\t_2(t)=\ww\rho(t)^{-1}\ww a(t)$. \index{If, in addition, the process $\ww\rho(t)$ is non-random, then  the process $\ww
  S(t)$ has the same distribution under $\P_\t$ for all $\t\in\T$.}
\item[(iii)] Benninga {\em et al.} (2002) considered a
multi-stock market with special requirements for the  equivalent martingale
measure.  For our special case of a single stock and a stochastic num\'eraire, these requirements leads to a unique  equivalent martingale measure
such that the process $(\ww S
(t),\exp\left(\int_0^tk(s)ds\right)B(t)^{-1})$ is a martingale, for a given process $k(t)\ge 0$.
\item[(iv)]
 Let  $z(t)=\int_0^t |V(s)|^{-1}V(s)^\top dW(s)$; by L\'evy's characterization theorem, it is an one-dimensional Wiener process. Let $q(t)=|V(s)|^{-1}\ww
a(t)$. By the assumptions, it is a bounded process. Let $\w z(t)=\int_0^tq(s)ds+z(t).$  We have that  $V(t)^\top dW(t)=|V(t)|dz(t)$ and
\baaa d\ww S(t)=\ww S(t)\bigl(\ww a(t)dt+|V(t)|dz(t)\bigr)=\ww S(t)|V(t)|d\w
z(t).\eaaa By Girsanov's Theorem, there is an  equivalent martingale measure $\w\P$
such that $\w z(t)$ a Wiener process under $\w\P$.  It follows that
the process $\ww S(t)$ is an  equivalent martingale under $\w\P$. This martingale
measure was studied in Cheng (1991).
\end{itemize}
}
\end{example}
\par
Let $\Y$ be the set of all $\FF$-adapted measurable processes with
values in $\R^2$ that are  square integrable on $[0,T]\times\O$ with
respect to $\ell_1\times \P_\t $, where $\ell_1$ is the Lebesgue
measure. \par Let $\H_\t$ be the Hilbert space formed as the
completion of the set of $\FF$-adapted measurable processes $y(t)$
such that $\|y\|_{\H_\t}=\left(\E_\t\int_0^T|\ww S(t)y(t)|^2dt\right)^{1/2}<+\infty$.
\subsection*{Wealth and  discounted wealth}
 Let $X(0)>0$ be the initial wealth at time $t=0$ and let $X(t)$ be
the wealth at time $t>0$.
\par
We assume that the wealth $X(t)$ at time $t\ge 0$ is
\begin{equation}
\label{X} X(t)=\b(t)B(t)+\g(t)S(t).
\end{equation}
Here $\b(t)$ is the quantity of the num\'eraire portfolio, $\g(t)$ is the
quantity of the stock  portfolio, $t\ge 0$. The pair $(\b(\cdot),
\g(\cdot))$ describes the state of the  securities
portfolio at time $t$. Each of  these pairs is  called a strategy.
\begin{definition}
\label{adm} Let $\t\in\T$ be given. A pair $(\b(\cdot),\g(\cdot))$  is said to be an
admissible strategy under $\P_\t$ if the processes $\b(t)$ and
$\g(t)$ are progressively measurable with respect to the filtration
$(\F_t)_{t\ge 0}$ and such that  \baa  \E_\t \int_0^{T} \ww
S(t)^2\g(t)^2dt<+\infty.\qquad \label{admiss}\eaa
\end{definition}
\begin{definition} Let $\t\in\T$ be given.
A pair $(\b(\cdot),\g(\cdot))$ that is  an
admissible strategy  under $\P_\t$ is said to be a self-financing
strategy, if there exists a sequence of Markov times
$\{T_k\}_{k=1}^{\infty}$ with respect to $(\F_t)_{t\ge 0}t$ such that $0\le
T_k\le T_{k+1}\le T$ for all $k$, $T_k\to T$ as $k\to +\infty$ a.s.,
and \baa &&\E_\t \int_0^{T_k}\left(\b(t)^2B(t)^2 +
S(t)^2\g(t)^2\right)dt<+\infty, \quad k=1,2,... \label{admiss0}\eaa
and the corresponding wealth $X(t)=\g(t)S(t)+\b(t)B(t)$ is such that
\begin{equation}
\label{self} dX(t)=\g(t)dS(t)+\b(t)dB(t).
\end{equation}
\end{definition}
Note that condition (\ref{admiss0}) ensures that the stochastic
differentials in (\ref{self}) are well defined.
\par

Let $\ww X(t)\defi X(t)/B(t)$. The process $ \ww X(t)$ is said to be
the discounted wealth. \par The following lemma is known; see, e.g., Geman
{\em et al.} (1995), Jamshidian (2008),  Bielecki {\em et al.} (2009).
\begin{lemma} \label{lemma1}
 If a strategy $(\b(t),\g(t))$ is
self-financing and admissible   under $\P_\t$ for some $\t\in\T$, then, for the corresponding discounted
wealth, \baa\label{CwwXev} d\ww X(t)=\g(t)d\ww S(t).\eaa
\end{lemma}
\begin{remark}
Since we assume that the coefficients for the equations for $S(t)$
and $B(t)$ are bounded, it follows from  Lemma
\ref{lemma1}  that if (\ref{admiss}) holds for some $\t$ then
$\E_\t \ww X(T)^2<+\infty$ for this $\t$.
\end{remark}
\begin{lemma}\label{lemmaRN} For every $\t\in\T$, the processes $\ww X(t)$
and $\ww S(t)$ are martingales under $\QQ$ with respect to $(\F_t)_{t\ge 0}t$,
i.e.,
  $\E_\t \{\ww S(T)\,|\F_t\}=\ww S(t)$ and  $\E_\t \{\ww X(T)\,|\F_t\}=\ww
  X(t)$.
\end{lemma}
\begin{remark}\label{anym} Consider an European option with the payoff $B(T)\xi$,
where $\xi$ is an $\F_T$-measurable random variable. For any
$\t\in\T$ such that $\E_\t\xi^2<+\infty$, the option price
$\E_\t\xi$ is an arbitrage-free price.
\end{remark}
\section{On the replicability   of contingent claims}\label{SecRep}
For $\t\in\T$, let $\X$ be the subspace of  $ L_2(\O,\F_T,\P_\t )$
consisting of all $\zeta\in L_2(\O,\F_T,\P_\t )$ such that there
exists an admissible self-financing strategy $(\b(\cdot),\g(\cdot))$
under $\P_\t$ and the corresponding  wealth process $X(t)$ such that
$X(0)=0$ and $X(T)=B(T)\zeta$. \par Let $\X^\bot\defi\{\eta\in
L_2(\O,\F_T,\P_\t ):\quad \E_\t \eta=0,\quad \E_\t [\zeta \eta]=0\
\hbox{for all}\quad  \zeta\in\X\}$.
\par
Let $\xi\in L_2(\O,\F_T,\P_\t )$. By the Martingale Representation
Theorem, we have that, for some uniquely defined $U_\t\in\Y$ and $c_\t\in\R$,  \baa
\xi=c_\t+\int_0^TU_\t(t)^\top dW_\t(t). \label{FSu}\eaa In addition,
it follows from the properties of closed subspaces in Hilbert spaces
that $\xi$ can be represented via F\"ollmer-Schweizer decomposition
\baa \xi=c_\t+I_\t+R_\t. \label{FS}\eaa Here $c_\t=\E_\t\xi$,
$R_\t\in\X^\bot$, and \baa I_\t=\int_0^T \g_\t(t)d\ww S(t)\in\X
\label{Ig}\eaa for some $\g_\t\in\H_\t$,  i.e., it is the terminal
discounted wealth $\ww X(T)$ for some admissible self-financing
strategy $(\b_\t(\cdot),\g_\t(\cdot))$ under $\P_\t$ and for the
initial wealth $X(0)=0$. Therefore, a contingent claim $B(T)\xi$
 can be decomposed as $B(T)(\ww\xi_\t+R_\t)$, where
$B(T)R_\t$ is the hedging error and where $B(T)\ww\xi_\t$ is a
replicable part such that $\ww\xi_\t=c_\t+\int_0^T \g_\t(t)d\ww
S(t)$.
\par
Let us express $\g_\t$ via $U_\t$.
\begin{proposition}\label{propA} Let
$\xi\in L_2(\O,\F_T,\P_\t )$, and let $U_\t$ be defined by (\ref{FSu}).
Let \baa \nuu_\t(t)=U_\t(t)^\top
V(t)/|V(t)|^2,\quad \eta_\t(t)=U_\t(t)-\nuu_\t(t)V(t).\label{eta}\eaa
Then \baaa
 \eta_\t(t)^\top V(t)\equiv 0,\eaaa
and (\ref{FS}) holds with  \baa
I_\t=\int_0^T\nuu_\t(t)V(t)^\top dW_\t(t), \quad
R_\t=\int_0^T\eta_\t(t)^\top dW_\t(t). \label{IR}\eaa Further,
(\ref{Ig}) holds with \baa \g_\t(t)=\nuu_\t(t)\ww
S(t)^{-1}.\label{g}\eaa
\end{proposition}
\par
{\it Proof.} It suffices to observe that $\eta_\t\in\Y$, and that $\nuu_\t V$ is the projection of $U_\t$ on $V$. It follows that $R_\t\in\X^\bot$.
The uniqueness follows from the properties of orthogonal subspaces
of a Hilbert space. $\Box$
\par
The following statement follows from the non-uniqueness of the
 equivalent martingale measures and The 2nd Fundamental Theorem of Asset
Pricing.
\begin{proposition}
\label{ThM} Assume  $\ww\rho(\cdot)\neq 0$, i.e., it is not an
identically zero process.  Then the set $\X^\bot$ contains non-zero
elements, i.e.,   $\sup_{\eta\in\X^\bot}\E_\t|\eta|>0$.
\end{proposition}
\par
By this proposition,  the hedging error $R_\t$ is non-zero in the
general case. In other word, a contingent claim of a general type is
not replicable. For completeness, we will give in the Appendix the
proof adjusted to our model.
\par
Let us describe some cases of replicability.
\par
Let $(\F_t^w)_{t\ge 0}$ be the filtration generated by the process
$w(t)$, and let $(\F_t^{\ww S})_{t\ge 0}$ be the filtration
generated by the process $\ww S(t)$.
\begin{theorem}\label{ThM00} Assume that the processes $\ww\s(t)$ and $\ww\rho(t)$ are non-random.
Then  the claims $B(T)\xi$ are replicable for  $\xi\in
L_2(\O,\F^{\ww S}_T,\P_\t )$  for any $\t\in\T$.  More precisely, there exists an $\FF$-adapted
process $\g(t)$ such that $\E_\t\int_0^T\g(t)^2\ww S(t)^2dt<+\infty$
   and $\xi=\E_\t\xi+\int_0^T\g(t)d\ww
S(t)$.
\end{theorem}
\subsubsection*{The case of complete market}
Note that the market described in Theorem \ref{ThM00}  is incomplete since there are claims that cannot be replicated.
The following theorem describes an important special case when the market is complete.
\begin{theorem}\label{ThM0} Assume that  the processes $\ww a(t)$ and  $\ww \s(t)$ are adapted
to the filtration $(\F_t^w)_{t\ge 0}$ generated by the process $w(t)$, and that $\ww\rho(t)\equiv 0$, i.e., it is  an
identically zero process. Then $\t_1(t)=\ww a(t)\ww\s(t)^{-1}$ for any $\t\in\T$, and  the claims $B(T)\xi$ are replicable for $\xi\in
L_2(\O,\F^w_T,\P_\t )$. \end{theorem}
\section{The implied market price of risk: the case of complete market}\label{SecImp}
Up to the end of this section, we assume that
that $\w\s(t)\equiv \w\rho(t)\equiv 0$.  In this case, by the definition of  $\F_t$, we have that  $\F_t=\F_t^w$, and the assumptions of Theorem \ref{ThM0}
are satisfied.
\begin{lemma}\label{corr2}
 The claim $\xi\equiv \$ 1$ is replicable in the following sense:
for any $t\in[0,T)$,   there exists an $\FF$-adapted
process $\g(t)$ such that $\E_\t\int_0^T\g(t)^2\ww S(t)^2dt<+\infty$
   and \baaa
   B(T)^{-1}=\E_\t B(T)^{-1}+\int_0^T\g(t)d\ww S(t).\eaaa
\end{lemma}
\par It can be noted that, by the definitions, under the assumption of Lemma \ref{corr2},
 the processes $a(t),\s(t),\a(t)$, and $\rho(t)$, are $\F_t^w$-adapted.
In addition, $\t_1(t)=\ww a(t)\ww\s(t)^{-1}$ for any $\t\in\T$.

Under the assumptions of Lemma \ref{corr2},
  the value $\E_\t B(T)^{-1}$ represents the price at time $t=0$ of a zero-coupon bond
with the payoff \$1 at the maturity time $T$. The value
 $\ww X(t)=\E_\t\left\{B(T)^{-1}|\F_t\right\}$ represents the discounted wealth for the
 replicating strategy, and the value \baa
 P(t,T)=B(t)\ww X(t)=B(t)\E_\t\left\{B(T)^{-1}|\F_t\right\}\label{bond}\eaa represents the total wealth for the
 replicating strategy and the price at time $t$ of a zero-coupon bond
with the payoff \$1 at the maturity time $T$.
\par
Let us discuss some consequences of these statements.
\par
Lemma \ref{corr2} implies that the value \baa
r(t)\defi -(T-t)^{-1}\log P(t,T)=-(T-t)^{-1}\log\left( B(t)\E_\t\left\{B(T)^{-1}|\F_t\right\}  \right)
\label{rbond}\eaa represents the so-called yield to maturity, or the expected average risk-free
rate associated with the zero-coupon bond, meaning that the price at time $t$ of a zero-coupon bond
with the payoff \$1 at the maturity time $T$ is
\baa
P(t,T)=\exp(-r(t)(T-t)).
\label{Pr}\eaa
\par
If the processes $a(t),\s(t),\a(t)$, $\rho(t)$, $\w\rho(t)$, $\t(t)$ are constant and
the assumptions of Lemma \ref{corr2} are satisfied,  then
\baa &&B(T)^{-1}=B(t)^{-1}\exp\left(\Bigl(-\a +\frac{\rho^2 }{2}+\rho\t_1\Bigr)[T-t]
-\rho \Bigl(W_{1\t}(T)-W_{1\t}(t)\Bigl)\right).\hphantom{xxx}\label{Binverse}
\eaa In this case, a direct calculation of (\ref{bond}) gives
\baaa
-(T-t)^{-1}\log P(t,T)=\a -\rho^2 -\rho\t_1. \eaaa
Since $\t_1=\ww a/\ww\s$, it gives that $
-(T-t)^{-1}\log P(t,T)=\a -\rho^2 -\rho\ww a/\ww\s. $
Hence  \baa
 r=\a -\rho^2-\rho\ww a/\ww\s\label{r}\eaa
  can be interpreted as the "true" risk-free rate for this market.
It can be seen that  $r$ is close to $\a$ if $\rho$ is small; if $\rho=0$ then $r=\a$.
\par
Further, let us  consider  a scenario where  the real market price $P_{\mar}(0,T)$ of a zero-coupon bond
with the payoff \$1 at the maturity time $T$ is observed from the market statistics  at time $t=0$, and the corresponding value  (\ref{bond})  $r_{\mar}=-T^{-1}\log P_{\mar}(0,T)$ is calculated.  If $\rho=0$ then only $r_{\mar}=\a$ is consistent
with (\ref{bond}). Assume that $\rho\neq 0$.  In this case, we can reverse pricing formula (\ref{bond})  and calculate {\em implied }
 $\t_{1,\imp}$ from (\ref{r}) as \baa\t_{1,\imp}=(\a -\rho^2-r_{\mar})/\rho.
 \label{imp1}
 \eaa
 \par
 In this case, equation (\ref{B}) can be rewritten as   \baaa
&&dB(t)=B(t)\bigl([r_{\mar}+\rho^2]dt+\rho dW_{1\t}(t)\bigr),\eaaa
where $W_{1\t}(t)=w(t)+\int_0^t\t_{1,\imp}(s)ds$.
\par
It particular, it follows that a choice of $\rho$ for a market model with given $\a$  is consistent with the observed bond prices
if  $(\a -r_{\mar})/\rho$ is bounded as $\rho\to 0$. This leads to the following heuristic rule: if the observed bond market price is such that $\a -r_{\mar}$ is large, then one should  assume a sufficiently large $\rho$, to avoid  overestimation of the market price of risk.
\par
Representation  (\ref{imp1})  follows the classical approach to the so-called implied volatility
where the Black-Scholes formula is reversed.  However, there is some novelty:
 as far as we know, this is the first  attempt to derive the implied market price of risk process generated by the appreciation rate of  a stock  prices.  Currently, there are few other implied processes considered in the literature, besides the classical implied volatility.
Turvey and  Komar (2006) considered inference of the implied value  $a-\t_1\s$  from the market option price,  in a model that corresponds to our model with $\w\s=\rho=\w\rho=0$,  presuming that this value is used as the appreciation rate under the pricing measure; for the Black and Scholes model, this value should be the risk-free rate. The implied cumulate  risk-free rate was considered in the framework of Black and Scholes model in Dokuchaev (2006) and Hin and Dokuchaev (2014),   as an inferred parameters from stock option prices. Weron (2008) estimated the implied market  price of risk for energy  prices as the difference between  the observable historical Ornstein-Uhlenbeck long term mean and the implied long-term mean infrared from the market  options prices.  Finally, the implied martingale measure for a bond market was introduced in Bielecki {\em et al.}  (2009); this construction was based on observation of bond prices for a continuum of maturities.  None of these papers considered estimation of the appreciation rate of the stock as an implied parameter inferred from the stock option  prices.
The  implied martingale measure defined by (\ref{imp1}) has a different nature: it is associated with a market price of risk process $\ww a(t)/\ww \s(t)$  and the appreciation rate $a(t)$ of the stock  prices.

It appears that estimation
of the appreciation rate of stock prices and the market price of risk process $\t(t)$ from the historical data is a quite challenging problem that is important for financial applications, especially  for optimal
 portfolio selection.  For financial models, estimation of these  processes  is more difficult than estimation of the volatility   since the trend for financial time series is usually relatively small and unstable. Some results and references for the estimation of the appreciation rate and application to portfolio selection can be found in Brennan (1998), Dokuchaev (2005), and Dokuchaev (2002), Ch.9, p.128.
Calculation of the  implied $\t$
form observed bond prices as described above could be a useful addition to the existing methods. Further development of this approach is presented in Section \ref{SecImp2} below.

  \begin{remark} \label{remBond}
For the bond pricing model with constant coefficients described above, the choice of $(\t_1,r)$ is independent on $T$. It follows that  a single  market price
$P_{\mar}(0,T)$ of a zero coupon bond for
one given maturity time $T$ defines uniquely the prices of similar bonds for all other maturity times $\oo T\neq T$ given that these prices are defined by (\ref{bond}). This is caused by the fact that this formula has to be applied with the same $\t_1$ leading to the same $r$ in (\ref{Pr}).   It can be noted that, for models with time variable coefficients of equations for $(B,S)$, the same approach  gives a time dependent $(\t_1(t),r(t))$, and the value (\ref{r}) defined for a maturity time $\oo T$ depends on $\oo T$.
\end{remark}\begin{remark}\label{corrU}
Under the assumptions of Theorems \ref{ThM00}--\ref{ThM0}, the choice
of $\g$ is unique, i.e., it is the same for all  $\t\in\T$ such that
$\E_\t\xi^2<+\infty$; the
expectation $\E_\t\xi$ is also the same for all these $\t$ . \end{remark}
\index{ The proofs of all results are given in the Appendix.}
\section{On relativity of the price and the hedging error}\label{secE}
The number $c_\t=\E_\t\xi$  is commonly regarded as the price of an
option with the payoff $B(T)\xi$. This price depends on the
selection of $\t$. The following theorems demonstrate that this
price can be selected quite arbitrarily even for the case of an
arbitrarily small stochastic deviations in (\ref{B}),  i.e., for
arbitrarily small processes $\rho(t)$ and $\w\rho(t)$. For instance,
we can select $\rho(t)\equiv \w\rho(t)=\e$ for an arbitrarily small
$\e>0$. This means that the presence of small deviations in
(\ref{B}) changes dramatically the properties of the market model.
\def\K{\kappa}
\par
We denote $x^+=\max(0,x)$ for $x\in\R$.
 \begin{theorem}\label{ThR1} Assume that
\baa
\essinf_{t,\o}|\ww\s(t,\o)\w\rho(t,\o)-\rho(t,\o)\ww\rho(t,\o)|
>0.
\label{det}\eaa
 Let $\K\in (0,+\infty)$ be given, and let
$\xi=B(T)^{-1}(\K-S(T))^+$. Then the following holds.
\begin{itemize}
\item[(i)]
 for any $\e>0$, there exists $\t\in\T$
such that $c_\t=\E_\t\xi \in[0,\e]$, and  \item[(ii)] For any $M>0$,
there exists $\t\in\T$
 such that $c_\t=\E_\t\xi \ge M$.
\end{itemize}
\end{theorem}
\begin{theorem}\label{ThR2} Assume that
 (\ref{det}) holds. Let
$\K\in (0,+\infty)$ be given, and let $\xi=B(T)^{-1}(S(T)-{\K})^+$. Then the
following holds.
\begin{itemize}
\item[(i)]
For any $\e>0$, there exists $\t\in\T$  such that $c_\t
=\E_\t\xi\in[0,\e]$, and
\item[(ii)]
 For any $\e>0$, there exists $\t\in\T$
 such that $c_\t=\E_\t\xi \in[ S(0)-\e,S(0)]$. \end{itemize}
\end{theorem}
\par
Consider a hedging strategy that replicates the claim
$B(T)(c_\t+I_\t)$, where $c_\t\in\R$ and $I_\t\in\X$ are such that
(\ref{FS}) holds with some $R_\t\in\X^\bot$. This $R_\t$ is the
hedging error.
\par
The following theorems show that the value of the second moment of
$R_\t$ is varying  widely  with variations of the historical measure
and take can extreme values for some choices of  the equivalent martingale measures.

\begin{theorem}\label{ThReli} Let $\xi$ be a random claim such that (\ref{FS}) holds for
 some $\t\in\T$,  $c_\t\in\R$, $I_\t\in\X$, and  $R_\t
 \in\X^\bot$ such that $\E_\t R_\t^2>0$. Assume that (\ref{FSu}) holds
 for $U_\t\in\Y$ such that \baaa
 \esssup_{\o}\int_0^T|U_\t(t,\o)|^2dt<+\infty,\qquad\essinf_{\o}\int_0^T|\eta_\t(t,\o)|dt>0\eaaa
for the process $\eta_\t$ defined by (\ref{eta}).   Then the following holds:
\begin{itemize}\item[(i)]
  For any $M>0$,  $\P(R_\t^2>M)>0$;
  \item[(ii)] For any $M>0$, there exists a measure $Q$ that is
equivalent to $\P$ and such that $\E_QR_\t^2\ge M$, where $\E_Q$
is the corresponding expectation.
\end{itemize}
\end{theorem}
\begin{theorem}\label{ThRelii} Let $\xi$ be a random claim such that (\ref{FS}) holds for
 some $\t\in\T$, $c_\t\in\R$, $I_\t\in\X$, and  $R_\t
 \in\X^\bot$ such that $\E_\t R_\t^2>0$. Assume that (\ref{FSu}) holds for
  $U_\t\in\Y$ such that \baaa
  \esssup_{t,\o}|U_\t(t,\o)|<+\infty,\qquad
  \essinf_{t,\o}|\eta_\t(t,\o)|>0\eaaa
 for the process $\eta_\t$ defined by (\ref{eta}).  Then
 \begin{itemize}\item[(i)]
  For any $\e>0$,  $\P(R_\t^2<\e)>0$;
  \item[(ii)]  for any  $\e>0$, there exists a measure $Q$ that is equivalent to $\P$
and such that $\E_QR_\t^2\le \e$, where $\E_Q$ is the
corresponding expectation.
\end{itemize}\end{theorem}
\begin{remark} Clearly, the statement of Theorem  \ref{ThReli} holds if and only if $\P(R_\t^2>M)>0$
for any $M>0$, and  the statement of Theorem  \ref{ThRelii} holds if and only if $\P(R_\t^2<\e)>0$
for any $\e>0$.  Therefore, the statements of Theorems  \ref{ThReli}--\ref{ThRelii} can be reformulated as the following: the assumptions  imposed there on $U$ and $\eta_\t$
imply the corresponding  conditions (i).
\end{remark}
\section{On selection of $\t$ and the  equivalent martingale measure}\label{secS}
Since the  equivalent martingale measure is not unique, a question arises which
particular $\t$ should be used for calculation of the price $c_\t=\E
_\t\xi$. In the literature,  there are many methods developed for
this problem, mainly for the incomplete market models with random
volatility and appreciation rate. \par
One may look for "optimal" $\t$ and $c_\t$ in the spirit of
mean-variance pricing, such that $\E R_\t^2$ is minimal; see, e.g.,
Schweizer (2001). A generalization of this approach leads to
minimization of $\E|R_\t|^q$ for $q\ge 1$. An alternative approach
is to define the price as $\sup_{\t\in\T_0}c_\t$ for some reasonably
selected set $\T_0\subset \T$.  In the case of an incomplete market
with random volatility, this pricing rule leads to a corrected
volatility smile (Dokuchaev (2011)).
\par
The following Theorem \ref{ThN}  will be useful for selection of $\t$.
\renewcommand{\arraystretch}{1.0}
\begin{theorem} \label{ThN}  Let $\t=(\t_1,\t_2)^\top\in\T $ be given,
and let  $\varrho(t)=\w V(t)^\top\t(t)$, i.e.,
\baa
&&\ww\s\t_1+\ww\rho\t_2=\ww a,\nonumber\\
&&\rho \t_1+\w\rho\t_2=\varrho. \label{sys}\
\eaa
Then
  \baa
  && dS(t)=S(t)\bigl([a(t)-\ww a(t)-\varrho(t)]dt+\s(t) dW_{1\t}(t) +\w\s(t)dW_{2\t}(t)\bigr),\nonumber\\
&&dB(t)=B(t)([\a(t)-\varrho(t)]dt+\rho(t) dW_{1\t}(t))+\w\rho(t) dW_{2\t}(t)).\label{SB}\eaa
\end{theorem}
\par
Clearly, if (\ref{det}) holds, then a unique $\t$ can be found from system (\ref{sys})  for any $\FF$-adapted and bounded process  $\varrho(t)$.
In this case, Theorem \ref{ThN} gives  a useful parametrization of the set $\T$ via $\varrho$.
Examples \ref{Ex1}-\ref{Ex4} below  demonstrate  how parametrization of $\T$ via $\varrho$ helps
  to find some reasonable choices of $\t$.  For these examples, we assume
that (\ref{det}) holds.
\begin{example}\label{Ex1} {\rm\par
 For $\t$ from Theorem \ref{ThN} with $\varrho\equiv 0$,  the process $(S(t),B(t))$ evolves as
 \baaa
  && dS(t)=S(t)\bigl([a(t)-\ww a(t)]dt+\s(t) dW_{1\t}(t) +\w\s(t)dW_{2\t}(t)\bigr),\\
&&dB(t)=B(t)\bigl(\a(t)dt+\rho(t) dW_{1\t}(t))+\w\rho(t) dW_{2\t}(t)\bigr).\label{SB1}\eaaa
In particular, the equation for $B$ has the same coefficients as the equation for $B(t)$ under $\P$,
with replacement of $W(t)$ by $W_\t(t)$. In particular, the
distribution of $B(t)$ under $\P_\t$ and under the historical
measure $\P$ is the same if the coefficients $\a(t)$, $\rho(t)$, and
$\w\rho(t)$, are non-random.
In addition, if $\w\s\equiv 0$ then the choice $\varrho\equiv 0$ ensures that $\t_1=\w a/\s$.     }
\end{example}
\begin{example}\label{Ex2}  {\rm For
$\t$ from Theorem \ref{ThN} with $\varrho=-\ww a $,  the evolution of $S$
under $\P_\t$  is described by an It\^o equation with the same
coefficients as the equation for $S(t)$ under $\P$, with replacement
of $W(t)$ by $W_\t(t)$. The distribution of $S(t)$ under $\P_\t$ and
under the historical measure $\P$ is the same for the case of
non-random coefficients $a(t)$, $\s(t)$, and $\w\s(t)$.
}\end{example}

\begin{example}\label{Ex3} {\rm   Let $k\in(0,1)$ be given. Let us
calculate $\t$ from (\ref{sys}) with $\varrho=\frac{k}{1-k}(\rho^2+\w\rho^2),$ i.e.,  \baaa
&&\ww\s\t_1+\ww\rho\t_2=\ww a,\nonumber\\
&&\rho \t_1+\w\rho\t_2=\frac{k}{1-k}(\rho^2+\w\rho^2).\label{ThetaBond}
\eaaa
 The second equation here can be rewritten as
 \baaa
&&\rho \t_1+\w\rho\t_2=k(\rho^2+\w\rho^2+\rho\t_1+\w\rho\t_2).
\eaaa
Let  $r_\t$ be defined by (\ref{rf}). This implies  that $
k(\rho^2+\w\rho^2+\rho\t_1+\w\rho\t_2)=k(\a-r_\t)$.  It follows that $\t$
satisfies system (\ref{sys}) with  $\varrho=\varrho(\t)=k(\a-r_\t)$.  By (\ref{SB}), this leads to the equation   \baaa
&&dB(t)=B(t)\bigl([kr_\t+(1-k)\a]dt+\rho dW_{1\t}(t))+\w\rho dW_{2\t}(t)\bigr),\eaaa
i.e., the appreciation rate coefficient
 for  $B$ under $\P_\t$  is $kr_\t+(1-k)\a$.
 Therefore, we have established that there exists a  choice of  $\t$  that ensures that
the appreciation rate
 for  $B$ under $\P_\t$  can be arbitrarily   close to  $r_\t$  representing  the expected average risk-free
rate associated with the zero-coupon bond under the measure $\P_\t$.
This can be achieved with selection of $k$ close to 1.
}\end{example}
\par
\begin{example}\label{Ex4} {\rm  An important example of the selection of $\t$ is
 \baa \t(t)=\ww a(t)V(t)/|V(t)|^2.\label{min}\eaa
}\end{example}
The following theorem shows that this corresponds to the choice of $\t$ with the minimal norm.
\begin{theorem} \label{ThLRM} Let
 $\t(t)$ be defined by (\ref{min}). Then, for
 every $t,\o$,
 the value of $|\t(t,\o)|$ is minimal among all  $\t\in\T$. In addition, if
 $\xi=c_\t+\int_0^T\g_\t(t) d\ww S(t)+R_\t$ for some   $R_\t\in\X^\top$ and
$\g_\t$ is an adapted process such that $\g_\t\s\in\H_\t$, then $\E
(R_\t\M(T))=0$, where $\M(T)=\int_0^T\g_\t(t)\ww S(t)V(t)^\top
dW(t)$ represents the "martingale" part of the integral \baaa
\int_0^T\g_\t(t)d\ww S(t)=\int_0^T\g_\t(t)\ww S(t)\w a(t)
dt+\M(T).\eaaa
\end{theorem}
\par
The selection of $\t$ described in Theorem \ref{ThLRM} ensures that
the corresponding self-financing strategy with the quantity of
shares $\g(t)$ is a so-called {\em locally risk minimizing
strategy}; see, e.g., F\"ollmer and Sondermann (1986), Biagini and
Pratelli  (1999).
\par
Let us reconsider Example \ref{exCheng}
(iv). We will be using  the measure $\w \P$ and the processes $q(t)$, $z(t)$, and $\w z(t)$ defined in this example.

Set  $\V(t) = \w V(t)-k(t)V$, where \baaa k(t)=\w
V(t)^\top V(t)/|V(t)|^2.\eaaa Clearly, we have that $\V(t)^\top V(t)=0$.

 Further,
there exists a one-dimensional Wiener process $z_1(t)$ such that
$\int_0^t \V(s)^\top dW(s)=\int_0^t|\V(s)|dz_1(s)$ and \baaa
dB(t)&=&B(t)\bigl(\a(t)dt+k(t)V(t)^\top
dW(t)+\V(t)^\top dW(t)\bigr)\\&=&B(t)\bigl(\a(t)dt+k(t)|V(t)|dz(t)+|\V(t)|dz_1(t)\bigr).\eaaa
For $q(t)=\ww a(t)/|V(t)|$, we have \baaa dB(t)=B(t)\bigl(\a(t)dt+k(t)
|V(t)|(d\w z(t)-q(t)dt)+|\V(t)|dz_1(t)\bigr).\eaaa On the other hand, \baaa dB(t)=B(t)\bigl(\a(t)dt+\w V(t)^\top
dW(t)\bigr)=B(t)\bigl(\a(t)dt+\w V(t)^\top(dW_\t(t)-\t(t)dt\bigr).\eaaa This means
that, in our notation, $\w\P=\P_\t$, where $\t\in\T$ is such that
\baaa k(t)q(t)|V(t)|=k(t)\ww a(t)=\w V(t)^\top \t(t). \eaaa The only
$\t\in\T$ satisfying this is $\t(t)=\ww a(t)V(t)/|V(t)|^2$ from
Theorem \ref{ThLRM}.
\section{The implied market price of risk for  an incomplete market}\label{SecImp2}
In Section \ref{SecImp}, we established that the value  (\ref{bond}) represents  the price at time $t$ of a zero-coupon bond
with the payoff \$1 at the maturity time $T$ for a case where
the claim \$1 is replicable.  This implies that it could be reasonable  to accept  (\ref{bond})
as the  price of this bond for a more general case of a non-replicable claim \$1 as well (i.e., with non-zero $\rho$ and $\w\rho$).
This means that $r(t)$ defined by   (\ref{rbond}) represents again the expected average risk-free
rate associated with the zero-coupon bond.

This would require to produce on more example of selection of $\t$.

Assume that  the processes
$a(t),\s(t),\a(t)$, $\rho(t)$, $\w\rho(t)$, $\t(t)$ are constant,
and that $r(t)$ is defined by (\ref{rbond}).
It follows from It\^o formula that
\baa B(T)^{-1}&=&B(t)^{-1}\exp\biggl(\Bigl(-\a
+\frac{\rho^2 }{2}+\frac{\w\rho^2 }{2}+\rho\t_1+\w\rho\t_2\Bigr)(T-t)\nonumber
\\&&-\rho \Bigl(W_{1\t}(T)-W_{1\t}(t)\Bigr)-
\w\rho \Bigl(W_{2\t}(T)-W_{2\t}(t)\Bigr)\biggr).\label{BinverseFull}
\eaa
 Then
(\ref{rbond})  imply that $r(t)$ can be found explicitly; it does not depend on $t$ and $T$ and depend on $\t$ such that $r(t)\equiv r_\t$, where  \baa r_\t=\a
-\rho^2-\w\rho^2-\rho\t_1-\w\rho\t_2\label{rf}.\eaa
\par
Equation (\ref{rf})   can be rewritten as
\baaa  \rho\t_1-\w\rho\t_2=\a- r_\t
-\rho^2-\w\rho^2. \label{rbond1}\eaaa
\par
Following the approach from Section \ref{SecImp}, consider now a scenario where  the real market price $P_{\mar}(0,T)$ of a zero-coupon bond
with the payoff \$1 at the maturity time $T$ is observed from the market statistics  at time $t=0$, and the corresponding value  (\ref{bond})  $r_{\mar}=-T^{-1}\log P_{\mar}(0,T)$ is calculated.  We can reverse pricing formula (\ref{bond})  and calculate {\em implied }
 $\t=\t_{\imp}$ as solution of  system (\ref{sys}) with $$\varrho=\a- r_{\mar}
-\rho^2-\w\rho^2.$$ This would follow again the classical approach to the so-called implied volatility
where the Black-Scholes formula is reversed.
\par
In particular,  equation (\ref{SB}) for this $\varrho$ implies that  \baaa
&&dB(t)=B(t)\bigl([r_{\mar}+\rho^2+\w\rho^2 ]dt+\rho dW_{1\t}(t))+\w\rho dW_{2\t}(t)\bigr),\eaaa
where $\t=\t_{\imp}$.
\par
This model has  the same feature as
described in Remark \ref{remBond} for a special model.  In particular, the choice of $\t$ is independent on $T$, and  a single  market price
$P_{\mar}(0,T)$ of a zero coupon bond for
one given maturity time $T$ defines uniquely the prices  defined by (\ref{bond}) for similar bonds for all other maturity times $\oo T\neq T$,
 since this formula has to be applied with the same $\t$.  For models with time variable coefficients of equations for $B$ and $S$, the same approach  gives a time dependent solution  $\t(t)$ of (\ref{sys}), and the value $r$ defined by (\ref{r}) for a maturity time $\oo T$ depends on $\oo T$.

\section{Markov case}\label{secM}
\def\ss{{\rm s}}
\def\bb{{\rm b}}
In this section,  we suggests some representations of prices, errors, and hedging strategies
via solutions of deterministic PDEs. \par
We will be using the processes ${\rm s}(t)=\log\ww
S(t)$ and ${\rm b}(t)=\log B(t)$. By It\^o formula, we obtain that
\baa &&d\ss(t)=(\ww a -\ww\s^2/2-\ww\rho^2/2)dt+\ww\s dw(t)+\ww\rho
d\w w(t),
\nonumber\\
&&d\bb(t)=(r -\rho^2/2-\w\rho^2/2)dt+\rho dw(t)+\w\rho d\w w(t).
\label{log}\eaa
\par
Assume that $\t\in \T$ is given. In this section, we will  assume
that there exists a measurable function $f: \R^2  \times [0,T]\to \R^8$
such that \baaa (\ww
a(t),\ww\s(t),\ww\rho(t),\a(t),\rho(t),\w\rho(t),\t_1(t),\t_2(t))^\top=
f(\ss(t),\bb(t),t).\eaaa  To simplify notation, we will describe it
as the following: we assume that there are measurable functions $\ww
a(s,b,t)$, $\ww\s(s,b,t)$, $\ww\rho(s,b,t)$, $\a(s,b,t)$, $\rho(s,b,t)$, $\w
\rho(s,b,t)$, $\t(s,b,t)$ defined on $\R^2\times[0,T]$ and such that
the processes $\ww a(t)$, $\ww\s(t)$,  $\ww\rho(t)$, $\a(t)$, $\rho(t)$,
$\w\rho(t)$, $\t(t)$ (defined on $[0,T]\times \O$) are replaced  by
the processes $\ww a(\ss(t),\bb(t),t)$, $\ww\s(\ss(t),\bb(t),t)$,  $\ww\rho(\ss(t),\bb(t),t)$,
$\a(\ss(t),\bb(t),t)$, $\rho(\ss(t),\bb(t),t)$,
$\w\rho(\ss(t),\bb(t),t)$, and $\t(\ss(t),\bb(t),t)$, respectively.
\par
Let  $\xi=F(\ww S(T),B(T))$, where $F:(0,+\infty)^2\to\R$ is a
measurable function  such that $\E_\t \xi^2<+\infty$ for some
$\t\in\T$. Consider the pricing and hedging problem for the claim
$B(T)\xi$. Let us calculate  the price $c_\t=\E_\t\xi$, and the hedging
strategy $\g(t)$ in (\ref{FS}) and (\ref{IR}).
\par
Let $H=H_\t=H_\t(s,b,t)$ be the solution of the following linear
parabolic equation in $\R^2\times[0,T]$: \baa &&H'_t+H'_s(\ww a
-\ww\s^2/2-{\ww\rho\,}^2/2) + H'_b (\a -\rho^2/2-{\w\rho\,}^2/2) + \L
H =H_s'\t_1+H_b'\t_2,\nonumber
\\
&&H(s,b,T)=F(e^s,e^b). \label{Eq} \eaa
\renewcommand{\arraystretch}{1.0} Here \baaa \hphantom{x}\L H=  \frac{1}{2} \left(
\begin{array}{c}
               \ww\s \\
               \rho
             \end{array}\right)^\top H''  \left( \begin{array}{c}
               \ww\s  \\
               \rho
             \end{array}\right)  + \frac{1}{2} \left(
\begin{array}{c}
                \ww\rho \\
                \w\rho
             \end{array}\right)^\top H''  \left( \begin{array}{c}
                \ww\rho \\
                \w\rho
             \end{array}\right),\quad\quad H''=\left(
\begin{array}{cc}
               H_{ss}'' & H_{sb}'' \\
               H_{bs}'' & H_{bb}''
             \end{array}\right).
             \eaaa

\renewcommand{\arraystretch}{1.4}             \par
             Assume that there exists a generalized solution $H(s,b,t)$ of (\ref{Eq})
             such that its gradient with respect to $(s,b)$ is bounded.
 By It\^o formula, it
follows that (\ref{FSu}) holds with \baaa
U_\t(t)=H'_s(\ss(t),\bb(t),t)V(t)+H'_b(\ss(t),\bb(t),t)\w V(t).\eaaa

In this case, (\ref{FS}) and (\ref{IR}) hold with \baaa c_\t=H(\log
S(0),\log B(0),0),\quad \g(t)=\nuu_\t(t)\ww S(t)^{-1}, \eaaa where
$\nuu_\t(t)=\frac{U_\t^\top V}{|V|^2}=f_\t(\ss(t),\bb(t),t)$, and where
\renewcommand{\arraystretch}{1.0}
          \baaa
&&f_\t(s,b,t)=H'_s(\ss(t),\bb(t),t)+\frac{H'_b(\ss(t),\bb(t),t)(\ww\s(s,b,t)\rho(s,b,t)+\ww\rho(s,b,t)\w\rho(s,b,t))}{\ww\s(s,b,t)^2+\ww\rho(s,b,t)^2}.
             \eaaa
             \index{OLD:
\renewcommand{\arraystretch}{1.0}
          \baaa
&&f_\t(s,b,t)=\frac{\ww\s(s,b,t)H'_s(s,b,t)-\w\rho(s,b,t)H'_b(s,b,t)}{\sqrt{\ww\s(s,b,t)^2+\ww\rho(s,b,t)^2}}
             \,\,.
             \eaaa}
\renewcommand{\arraystretch}{1.4}
\par
Further, let us consider the problem of calculation of $\E R_\t^2$,
i.e., estimation of  the hedging error with respect to the
historical measure $\P$. We have that (\ref{FS})-({\ref{IR}) hold
with $\eta_\t(t)=U_\t(t)-\nuu_\t(t)V(t)=g_\t(\ss(t),\bb(t),t)$, where
\renewcommand{\arraystretch}{1.0}
          \baaa
g_\t(s,b,t)= H'_s(s,b,t)V(t)+H'_b(s,b,t)\w V(t) -f_\t(s,b,t)\left(
            \begin{array}{c}
              \ww\s(s,b,t) \\
               \ww\rho(s,b,t)
            \end{array}
          \right)
      . \eaaa
\par
Let \baaa &&A(s,b,t)=g_\t(s,b,t)^\top \t(s,b,t),\quad
v(s,b,t)=g_\t(s,b,t).\eaaa

\par
It can be noted that, under the assumption of Proposition
\ref{propA}, $R_\t=x(T)$, where the process $x(t)$ evolves as \baa
&& dx(t)=\eta_\t(t)^\top\t(t)dt+\eta_\t(t)^\top dW(t), \quad x(0)=0.
\label{xforR}\eaa
By (\ref{xforR}), $\E R_\t^2=\E x(T)^2$, where \baa &&
dx(t)=A(\ss(t),\bb(t),t)dt+v(\ss(t),\bb(t),t)^\top dW(t).
\label{FAsOlog}\eaa This equation together with (\ref{log})
describes evolution of a diffusion Markov process
$(x(t),\ss(t),\bb(t))$. Therefore,
\baa \E   R_\t^2=\E x(T)^2=J(0,\log
S(0),\log B(0),0),\eaa where $J(x,s,b,t)$ is the solution of the
corresponding backward Kolmogorov-Fokker-Planck parabolic equation
for $(x(t),\ss(t),\bb(t))$ in $\R^3\times [0,T]$
 \baa
&&J'_t+J_x'A+  J_s' (\ww a -\ww\s^2/2-\ww\rho^2/2) +  J_b'(r
-\rho^2/2-\w\rho^2/2)+{\cal D}J,\nonumber
\\
&&J(x,y,z,T)=x^2. \label{FBelEq}\eaa Here \baaa {\cal D}J=
\frac{1}{2} \left(
\begin{array}{c}v_1\\
               \ww\s \\
               \rho
             \end{array}\right)^\top J''  \left( \begin{array}{c}
              v_1\\ \ww\s  \\
               \rho
             \end{array}\right)  + \frac{1}{2} \left(
\begin{array}{c} v_2\\
                \ww\rho \\
                \w\rho
             \end{array}\right)^\top J''  \left( \begin{array}{c}
               v_2\\ \ww\rho \\
                \w\rho
             \end{array}\right),\qquad J''=\left(
\begin{array}{ccc}
               J_{xx}'' & J_{xy}'' & J_{xz}''\\
               J_{yx}'' & J_{yy}'' & J_{yz}''\\
             J_{zx}'' & J_{zy}'' & J_{zz}''
              \end{array}\right).
             \eaaa
\setcounter{equation}{0}
\section{Conclusions} \par
We revisited the problem of pricing of stock options for the case of
the market model with a stochastic num\'eraire, with
emphasize on the impact of the non-uniqueness of  equivalent martingale measures
and the presence of non-replicable claims.  We found that there are
some interesting features that makes this model different from
incomplete market models where incompleteness is caused by the
randomness of the volatility. We found that the martingale prices
vary significantly for some extreme choices of the  equivalent martingale
measures. For instance, for a European put option, any sufficiently
large positive real number is a martingale price for some  equivalent martingale
measure. Some possible economically justified choices of  equivalent martingale
measures are suggested,  including a measure that correspond to a
consensus about the future num\'eraire process,  a measure that ensures local risk
minimizing hedging strategy, and  an implied measure that
takes into account observed zero coupon bond prices. This last measure is especially interesting,
since it leads to a formula for an implied market price of risk.
\par
It could be interesting to
consider optimal selection of the  equivalent martingale measure in the spirit
of the mean-variance hedging. It also could be interesting to develop a comprehensive bond pricing
model that is based on stochastic num\'eraire with time dependent coefficients, and investigate dependence of the price on the maturity time.  We leave it for future research.
 \section*{Appendix:  Proofs}
 \par
{\em Proof of Lemma \ref{lemma1}}  is straightforward and based on the application of It\^o's formula.
In fact,  Lemma \ref{lemma1}} represents a special case of Proposition 1 from Geman {\em et al.} (1995).  $\Box$
\setcounter{equation}{0}
\renewcommand{\theequation}{A.\arabic{equation}}
\renewcommand{\thelemma}{A.\arabic{lemma}}
\renewcommand{\theproposition}{A.\arabic{proposition}}
\par
{\em Proof of Lemma \ref{lemmaRN}} follows immediately from equation
(\ref{CwwXev})   and from the fact that $d\ww S(t)=\ww S(t)V(t)^\top
dW_\t (t)$. $\Box$
\par
{ \em Proof
 of Proposition \ref{ThM}.}
By Lemma \ref{lemma1} and \ref{lemmaRN}, the set $\X$ contains
random variables \baaa \int_0^T\g(t)d\ww S(t)=\int_0^T\g(t)\ww
S(t)V(t)^\top dW_\t (t), \eaaa where $\g\in\H_\t$ and where $W_\t$
is defined by (\ref{W}).
\par
For any $\zeta\in\X^\bot$, there exists $U(t)=(U_1(t),U_2(t))^\top
\in\Y$ such that \baaa \zeta=\int_0^TU(t)^\top dW_\t (t). \eaaa Let
us show that if $\zeta\in \X^\bot$ then $U(t)^\top V(t)=0$. For this
$\zeta$, we have that \baaa \E_\t \zeta\int_0^T\g(t)d\ww S(t) =\E_\t
\int_0^T\g(t)\ww S(t)V(t)^\top U(t)dt=0\quad \forall \g\in\H_\t.
\eaaa Hence $\ww S(t)V(t)^\top U(t)=0$ a.e. Hence $V(t)^\top U(t)=0$
a.e.
\par
  To show that the set
$\X^\bot$ contains non-zero elements, it suffices to take
$U_1(t)=\psi(t)\w\rho(t)$ and $U_2(t)=\psi(t)\ww\s(t)$, with an
arbitrary $\psi\in\Y$, i.e., \baa
\zeta=\int_0^T\psi(t)[\w\rho(t)dW_{\t 1 }(t)+\ww\s(t)dW_{\t 2}(t)].
\label{psi}\eaa This completes the proof. $\Box$
\par
{\em Proof of Theorem \ref{ThM00}}. Under the assumptions,  $d\ww
S(t)=\ww S(t)|V(t)|dz_\t(t)$, where $z_\t(t)$ is  a one-dimensional Wiener process such that $\int_0^tV(s)^\top dW_\t(s)=\int_0^t|V(s)|dz_\t(s)$.
Hence the filtration $(\F^{z_\t}_t)_{t\ge o}$ generated by $z_\t(t)$ is such that $\F^{z_\t}_T= \F^{\ww S}_T$.
Hence any $\xi\in
L_2(\O,\F^{\ww S}_T,\P_\t )$  belongs to
$L_2(\O,\F^{z_\t}_T,\P_\t )$.  By the Martingale
Representation Theorem, it follows that there exists an $\F^{z_\t}_t$-adapted
process $u_\t(t)$ such that $\E_\t\int_0^Tu_\t(t)^2dt<+\infty$
   and $\xi=\E_\t\xi+\int_0^Tu_\t(t)dz_\t(t)$. It suffices to select $\g(t)=u_\t(t)\ww S(t)^{-1}$. This completes
the proof of Theorem \ref{ThM00}.
$\Box$
\par
 {\em Proof
 of Theorem \ref{ThM0}}  follows immediately from the Martingale
Representation Theorem applied on the probability space $(\O,\F_T^w,\P_\t)$.
$\Box$
\par
{\em Proof of Lemma \ref{corr2}} follows from the assumptions of
Theorem \ref{ThM0} and from the equation  \baaa
B(T)^{-1}=B(t)^{-1}\exp\left([-\a +\rho^2 /2+\t_1](T-t)-\rho
[W_{1\t}(T)-W_{1\t}(t)]\right). \eaaa $\Box$

\par
{\em Proof of Remark \ref{corrU}}. Let the initial wealth $c_{\t
i}=X^{(i)}(0)$ and the strategy $(\b^{(i)}(\cdot),\g^{(i)}(\cdot))$
be such that $\ww X^{(i)}(T)=\xi$ a.s. for the corresponding
discounted wealth $X^{(i)}(t)$, $i=1,2$.
\par
 Set
$$ g(t)\defi \g^{(1)}(t)-\g^{(2)}(t),\qquad Y(t)\defi \ww
X^{(1)}(t)-\ww X^{(2)}(t). $$ We have  that $Y(T)=0$ a.s.. Hence
\baaa Y(T)=Y(0)+ \int_0^Tg(t)d\ww S(t)=0. \label{C2.44} \eaaa For
$K>0$, consider first exit times  $T_K=T\land\inf\{t: \int_0^t
(|\g^{(1)}(s)|+|\g^{(2)}(s)|^2ds\ge K\}$; they are Markov times with
respect to $(\F_t)_{t\ge 0}t$. We have that \baaa Y(T_K)=Y(0)+
\int_0^{T_K}g(t)d\ww S(t)=\E_{\t i}\{Y(T)\,|\,\F_{T_K}\}=0,\quad
i=1,2. \label{C2} \eaaa Hence \baaa
0=Y(0)^2+\E_{\t_i}\int_0^{T_K}g(s)^2\ww S(s)^2|V(s)|^2dt=0. \eaaa It
follows that $Y(0)=0$, and $g(t)|_{[0,T_K]}=0$ for any $K>0$. In
addition, $T_K\to T$ a.s. as $T_K\to +\infty$. Hence $g=0$. This
completes the proof of Remark \ref{corrU}. $\Box$

{\em Proof of Theorem \ref{ThR1}}. For any $K\in\R$, there exists
$\t=\t_K\in\T$ such that \baaa &&\t_1\ww\s+\t_2\ww\rho=\ww a\\
&& \t_1\rho+\t_2\w\rho=\w V(t)^\top \t(t)=K-r+\rho^2+\w\rho^2.\eaaa
By Girsanov's Theorem,
\renewcommand{\arraystretch}{1.0}\baaa W_\t(t)=
  W(t)+\int_0^t\t(s)ds.\label{W1}
\eaaa \renewcommand{\arraystretch}{1.4} is a standard Wiener process
in $\R^2$ under $\P_\t$.  We have that $d\ww S(t)=\ww S(t)V(t)^\top
dW_\t (t)$ and \baaa dB^{-1}(t)&=&B(t)^{-1}\bigl([-r+\rho^2+\w\rho^2
]dt-\rho dw(t)-\w\rho d\w
w(t)\bigr)\\&=&B(t)^{-1}\bigl([-r+\rho^2+\w\rho^2]dt-\w V(t)^\top
dW(t)\bigr)\\&=&B(t)^{-1}\bigl([-r+\rho^2+\w\rho^2]dt-\w V(t)^\top \t(t) dt+\w V(t)^\top dW_\t(t)\bigr)\\
&=&B(t)\bigl([-r+\rho^2+\w\rho^2]dt- (K-r+\rho^2+\w\rho^2)dt+\w V(t)^\top
dW_\t(t)\bigr)\\&=&B(t)^{-1}\bigl(-Kdt+\w V(t)^\top dW_\t(t)\bigr). \eaaa Let $\w B(t)^{-1}=e^{Kt}B(t)^{-1}$.
We have that \baaa && d\w B^{-1}(t)=\w B(t)^{-1}\w V(t)^\top dW_\t(t). \eaaa
 It
follows that $\w B(t)^{-1}$ is a martingale under
$\P_\t$.
\par
Let us prove statement (i). Let $K>0$. We have that \baaa \E_\t\xi=\E_\t
B(T)^{-1}(\K-S(T))^+&\le& \E_\t B(T)^{-1}\K=e^{-KT}\K\E_\t\w
B(T)^{-1}\\&=&e^{-KT}\K\w B(0)^{-1}\to 0 \quad\hbox{as} \quad{K\to
+\infty}. \eaaa
\par
Let us prove statement (ii).  Let $K<0$. We have
\baaa
\E_\t\xi&=&\E_\t B(T)^{-1}(\K-S(T))^+=\E_\t (B(T)^{-1}\K-\ww S(T))^+
\ge \E_\t B(T)^{-1}\K-E_\t \ww S(T)\\&=& \K  e^{-KT} \E_\t \w B(T)^{-1}-\ww S(0)=
 \K  e^{-KT} \w B(0)^{-1}-\ww S(0) \to
+\infty\quad\hbox{as} \quad{K\to -\infty}. \eaaa
 This completes the proof of Theorem \ref{ThR1}.
$\Box$
\par
{\em Proof of Theorem \ref{ThR2}}. Let  $\t=\t_K$ and $\w
B(t)$ be such as defined in the proof of Theorem \ref{ThR1}.
\par Let us prove statement (i).
We have that   \baaa \E_\t\xi=\E_\t
B(T)^{-1}(S(T)-\K)^+ =\E_\t
B(T)^{-1}(S(T)-\K) +\E_\t
B(T)^{-1}\Ind_{\{S(T)\le\K\}}(\K-S(T))
\\
\le \E_\t B(T)^{-1}S(T) -  \E_\t \Ind_{\{S(T)<\K\}}B(T)^{-1}S(T)\\
\le \E_\t B(T)^{-1}S(T) - \E_\t B(T)^{-1}S(T) +\E_\t \Ind_{\{S(T)>\K\}}B(T)^{-1}S(T)\\
\le \E_\t \Ind_{\{S(T)>\K\}}B(T)^{-1}S(T).
\eaaa
Let $\w S(t)=e^{Kt}S(t)$. By the definitions, we have that  \baaa dS(t)&=&S(t)\bigl(adt+V^\top dW(t)+
\w V^\top d W(t)\bigr)\\&=&S(t)\bigl(adt-V^\top \t dt -\w V^\top\t d t
 +V^\top dW_\t(t)+\w V^\top d W_\t(t)\bigr)\\&=&S(t)\bigl((a-\ww a)dt-
 (K-r+\rho+\w\rho^2)dt +V^\top dW_\t(t)+\w V^\top d W_\t(t)\bigr) \eaaa
 and
\baaa && d\w S(t)=\w S(t)\bigl((a-\ww a)dt+(r-\rho-\w\rho^2)dt +V^\top dW_\t(t)+\w V^\top d W_\t(t)\bigr). \eaaa
It follows
from the standard estimates for stochastic differential equations
that \baaa \sup_K\E_{\t_K}|\w S(T)|<+\infty, \eaaa  for $\t=\t_K$; see, e.g., Chapter 2 in
Krylov (1980). Hence  $\Ind_{\{S(T)>\K\}}=\Ind_{\{\w S(T)>\K e^{KT}\}}\to 0$ a.s. as $K\to +\infty$.
By the Lebesgue Dominated Convergence Theorem,  it follows  that  $\E_\t\xi\to 0$  as $K\to
+\infty$. Hence statement (i) follows.

\par
Let us prove statement (ii). For $K>0$, we have that \baaa \E_\t\xi&=&\E_\t
B(T)^{-1}(S(T)-\K)^+=\E_\t (\ww S(T)-B(T)^{-1}\K)^+\\&
\ge &\E_\t \ww S(T)-e^{-KT}\E_\t \w B(T)^{-1}
=\ww S(0)-e^{-KT}\w B(0)^{-1}\to S(0)\quad\hbox{as} \quad{K\to +\infty}. \eaaa
\par
In addition, we have that \baaa \E_\t\xi=\E_\t
B(T)^{-1}(S(T)-\K)^+=\E_\t (\ww S(T)-B(T)^{-1}\K)^+ \le \E_\t \ww
S(T)=S(0).\eaaa This completes the proof of Theorem \ref{ThR2}.
$\Box$
 \par
{\em Proof of Theorem \ref{ThReli}}. Since the statements (i) and (ii) are equivalent, it suffices to show that
(ii) holds.
\par
  Let $\sign(x)$ be defined such
that $\sign(x)=1$ for $x>0$, $\sign(x)=0$ for $x=0$, and
$\sign(x)=-1$ for $x<0$.
\par
 Let
$\psi(t)=(\sign(\eta_{\t 1}(s)),\sign(\eta_{\t 2}(s)))^\top$.
\par
 Let $K>0$, and  let measure $Q=Q_K$
be selected such that $$ W_Q(t)=W_\t(t)-K\int_0^t\psi(s)ds$$ is a
Wiener process under $Q$. This measure exists by  Girsanov's Theorem.
We have that \baaa R_\t=\int_0^T\eta_\t(t)^\top
dW_\t(t)=K\int_0^T\eta_\t(t)^\top\psi(t)dt +\int_0^T\eta_\t(t)^\top
dW_Q(t). \eaaa  Let \baaa
&&N_1(K)=\E_Q\left(\int_0^T\eta_\t(t)^\top\psi(t)dt\right)^2= \E_Q\left(\int_0^T(|\eta_{\t 1}(t)|+|\eta_{\t 2}(t)|)dt\right)^2,\\
&&N_2(K)=\E_Q\left(\int_0^T\eta_\t(t)^\top dW_Q(t)\right)^2. \eaaa We
have that \baaa
\inf_{K>0}N_1(K)=\inf_{K>0}\E_Q\left(\int_0^T(|\eta_{\t 1}(t)|+|\eta_{\t 2}(t)|)dt\right)^2
\ge\essinf_{\o}\left(\int_0^T|\eta_\t(t,\o)|dt\right)^2>0.
\eaaa
 By the definition of $\eta_\t$
and by assumptions on $U_\t$, it follows that
$\esssup_{t,\o}\int_0^T|\eta_\t(t,\o)|<+\infty$.
 Hence
 \baaa  \sup_{K>0}N_2(K)=\sup_{K>0}\E_Q \int_0^T|\eta_\t(t)|^2dt\le
\esssup_\o \int_0^T|\eta_\t(t,\o)|^2dt<+\infty. \eaaa Hence
$\E_QR^2_\t\ge K^2N_1(K)-2K\sqrt{N_1(K)N_2(K)}+N_2(K)\to +\infty$ as
$K\to +\infty$. This completes the proof of Theorem \ref{ThReli}.
$\Box$
\par
{\em Proof of Theorem \ref{ThRelii}}.  The statements (i) and (ii) are equivalent; it suffices to show that
(ii) holds.   Let $K>1$, and  let $y(t)$
evolves as \baaa dy(t)=\eta_\t(t)^\top dW_\t(t),\quad y(0)=0.
\eaaa Let $T_K=T\land\inf\{t>0:\,\, \int_0^ty(s)^2ds\ge  K\}$. Let
$q(t)=q_K(t)=(q_1(t),q_2(t))^\top$ be an $\FF$-adapted process such
that $q(t)^\top V(t)=0$, $q(t)^\top \eta_\t(t)=-Ky(t)$ for $t\le
T_K$, and $q(t)=0$ for $t> T_K$. By the assumptions on $\eta_\t$, it
follows that there is a unique process $q$ with this properties. In
addition, it follows that $|q(t)|\le C y(t)$  and
\baaa\int_0^Tq(s)^2ds=\int_0^{T_K}q(s)^2ds\le C\int_0^{T_K}y(s)^2ds\le CK,\eaaa where $C>0$ is defined by
$\essinf_{t,\o}|\eta_\t(t,\o)|$. \par
 Let a measure $Q=Q_K$
be selected such that $$ W_Q(t)=W_\t(t)-\int_0^t q(s)ds$$ is a
Wiener process under $Q$.  By the definitions, it follows that
\baaa dy(t)=-Ky(t)dt+\eta_\t(t)^\top dW_Q(t),\quad y(0)=0.
\eaaa
 By Girsanov's Theorem again, this measure
exists and is equivalent to $\P_\t$, and,  therefore, is equivalent
to $\P$. We have that \baaa R_\t&=&\int_0^T\eta_\t(t)^\top
dW_\t(t)=\int_0^T[\eta_\t(t)^\top q(t)dt +\eta_\t(t)^\top
dW_Q(t)]\\&=&-\int_0^{T_K}K y(t)dt +\int_0^{T}\eta_\t(t)^\top
dW_Q(t)=y(T_K)+ \int_{T_K}^{T}\eta_\t(t)^\top dW_Q(t).\eaaa

 \def\Ceta{C_\eta}
 By the assumptions on $U_\t$,
it follows that $\Ceta\defi \esssup_{t,\o}|\eta_\t(t,\o)|<+\infty$.
\par
 Clearly,
\baa
 &&\E_Qy(T_K)^2=\E_Q\int_0^{T_K}e^{-2K(T_K-s)}|\eta_\t(s)|^2ds\le \Ceta^2
\E_Q\int_0^{T_K}e^{-2K(T_K-s)}ds\nonumber\\&&= \Ceta^2
\E_Q\frac{1-e^{-2KT_K}}{2K}\le
\frac{1-e^{-2KT}}{2K}
 \to 0\quad \hbox{as}\quad K\to +\infty.\label{m1} \eaa
 \par
Consider events $A_K=\{\int_0^Ty(t)^2dt> K\}=\{T_K<T\}$.
 We have
\baaa \E_Q\left(\int_{T_K}^T\eta_\t(t)^\top dW_Q(t)\right)^2\le
\E_Q\int_{T_K}^T|\eta_\t(t)|^2dt\le
E_Q\Ind_{A_K}\int_{0}^T|\eta_\t(t)|^2dt\le  T\Ceta^2 Q(A_K). \eaaa
We have that
\baaa \E_Q y(t)^2dt \le \Ceta^2 \int_0^t e^{-2K(t-s)}ds \le\Ceta^2 T. \eaaa
By the Markov Inequality, it follows that \baa Q(A_K)&\le&
\frac{1}{ K}\E_Q\int_0^Ty(t)^2dt\le \frac{1}{ K} \Ceta^2 T^2
 \to 0\quad \hbox{as}\quad K\to +\infty.\label{m2} \eaa
 By (\ref{m1})-(\ref{m2}), $\E_QR_\t^2\to 0$ as $K\to +\infty$.
This completes the proof of Theorem \ref{ThRelii}. $\Box$
\par
{\em Proof of Theorem \ref{ThN}}.
 It can be verified directly that the equations for $S$ and $B$ have the desired form.   $\Box$
 \par
{\em Proof of Theorem \ref{ThLRM}}.  First, the standard Lagrange optimization techniques gives immediately that
  the selected  $\t$ is such that
$|\t(t,\o)|$ is minimal over all $\t\in\T$ and it is a unique solution of the problem \baaa
\hbox{Minimize}\quad |\t| \quad \hbox{subject to}\quad  V(t,\o)^\top
\t=\ww a(t,\o).\eaaa  Further, by Martingale
Representation Theorem, we have that, for some $U_\t\in\Y$, presentation
(\ref{FSu}) holds. It was shown in Section \ref{SecRep} that  (\ref{Ig})--(\ref{eta}) holds. By  Proposition \ref{propA}, we have that  $V(t)^\top
\eta_\t(t)\equiv 0$. For our choice of $\t$, this gives that  $\t(t)^\top
\eta_\t(t)\equiv 0$. It follows that \baaa
R_\t=\int_0^T\eta_\t(t)^\top dW_\t(t)=\int_0^T\eta_\t(t)^\top dW(t).
\eaaa Hence \baaa \E R_\t \int_0^T\g(t)\ww S(t)V(t)^\top dW(t)&=&\E
\int_0^T\eta_\t(t)^\top dW(t) \int_0^T\g(t)\ww S(t)V(t)^\top
dW(t)\\&=&\E\int_0^T\g(t)\ww S(t)\eta_\t(t)^\top V(t) dt=0.\eaaa This completes the
proof of Theorem \ref{ThLRM}. $\Box$
\subsection*{Acknowledgment} This work  was
supported by ARC grant of Australia DP120100928 to the
author.\section*{ References}$\hphantom{xx}$
K.  Back,  Martingale pricing,  {\em Annual Review of
Financial Economics}  {\bf 2} (2010)  235-250.

D. Becherer, The numeraire portfolio for unbounded semimartingales, {\em Finance and Stochastics} {\bf 5} (3) (2010) 327--341.

S. Benninga, T. Bjork, and
Z. Wiener,  On the use of numeraires in option pricing. {\em
Journal of Derivatives} {\bf 10} (2) (2002), 43--58. \index{Available at SSRN:
http://ssrn.com/abstract=297056 or doi:10.2139/ssrn.297056}

F. Biagini and M. Pratelli,  Local risk minimization and
numeraire. {\em Journal of Applied Probability}  {\bf  36}(4) (1999)
1126--1139.

T.R. Bielecki, M. Jeanblanc,  and M. Rutkowski,   Credit Risk Modeling (Osaka University Press, Japan, 2009).

M.J. Brennan,  { The role of learning in
dynamic portfolio decisions,}  {\em European Finance Review} {\bf 1} (1998),
295--306.

S.T.  Cheng,   On the feasibility of arbitrage-based option pricing
when stochastic bond price processes are involved,  {\em Journal of
Economic Theory}  {\bf 53}(1) (1991) 185--198.

N.  Dokuchaev,  Two
unconditionally implied parameters and volatility smiles and skews.
 {\it Applied Financial Economics Letters}  (2006)  {\bf 2}  199--204.

N.G. Dokuchaev, Optimal solution of investment
problems via linear parabolic equations generated by Kalman filter.
{\it SIAM J. of Control and Optimization} {\bf 44} (4) (2005)
1239--1258.

N. Dokuchaev,  {\it Mathematical finance: core theory, problems, and
statistical  algorithms} (Routledge,  New York, 2007).

N. Dokuchaev,  Option pricing via maximization over
uncertainty and correction of volatility smile,  {\it International
Journal of Theoretical and Applied Finance (IJTAF)}  {\bf 14} (4) (2011)  507--524.

N. El Karoui and M. Quenez,  Dynamic programming
and pricing of contingent claims in an incomplete market, {\em SIAM Journal on Control and
Optimization}  {\bf 33} (1) (1995) 29--66.

H. F\"ollmer and D.  Sondermann,    Hedging of non-redundant
contingent claims. In: W. Hildenbrand and A. Mas-Colell (eds.),
Contribution to Mathematical Economics (North Holland, New York, 1986) 205--223.

H.  Geman,  N. El Karoui,  and J.C. Rochet,
Changes of num\'{e}raire, changes of probability measure and option
pricing. {\em Journal of Applied Probability}  {\bf 32}(2) (1995),
443--458.

F. Jamshidian,   Numeraire invariance and application to
option pricing and hedging, working paper.
http://mpra.ub.uni-muenchen.de/7167/, 2008

I. Karatzas and C. Kardaras,   The num\'{e}raire portfolio in semimartingale financial models, {\em Finance and Stochastics}  {\bf 11}  (2007), 447--493.

C. Kardaras,   Num\'{e}raire-invariant preferences in financial modeling,
{\em  Ann. Appl. Probab.}  {\bf 20} (2010), 1697--1728.

Lin Yee Hin and N. Dokuchaev,   On the implied volatility layers under the future risk-free rate uncertainty,  {\em International Journal of Financial Markets and Derivatives} (2014), in press.

Yong-Jin  Kim and Naoto Kunitomo,  (1999).
  Pricing options under stochastic
interest rates: A New Approach. {\em Asia-Pacific Financial Markets}
{\bf 6} (1)  (1999)  49--70.

N.V. Krylov,   {\em Controlled Diffusion Processes}.\
(Springer, New York, 1980).

M. Schweizer,    A guided tour through quadratic hedging
approaches, {\em In: E. Jouini, J. Cvitanic, M. Musiela (eds.), "Option
Pricing, Interest Rates and Risk Management"}, Cambridge University
Press, 2001, 538--574.

C. G. Turvey and S. Komar,
 Martingale Restrictions and the Implied Market
Price of Risk, {\em Canadian Journal of Agricultural Economics} {\bf 54} (2006) 379--399.

R. Weron, Market price of risk implied by Asian-style electricity options and futures, {\em Energy Economics} {\bf 30} (2008) 1098--1115.

\end{document}